\documentclass[12pt]{article}
\usepackage{amsfonts}
\usepackage{amssymb}
\usepackage{graphics,psboxit,amsmath}

\def\hybrid{\topmargin 0pt    \oddsidemargin 0pt
        \headheight 0pt \headsep 0pt
        \textwidth 6.35in       
        \textheight 8.80in       
        \marginparwidth .875in
        \parskip 5pt plus 1pt   \jot = 1.5ex}

\hybrid

\def\baselinestretch{1.2}

\catcode`\@=11

\def\marginnote#1{}
\newcount\hour
\newcount\minute
\newtoks\amorpm
\hour=\time\divide\hour by60
\minute=\time{\multiply\hour by60 \global\advance\minute by-\hour}
\edef\standardtime{{\ifnum\hour<12 \global\amorpm={am}%
        \else\global\amorpm={pm}\advance\hour by-12 \fi
        \ifnum\hour=0 \hour=12 \fi
        \number\hour:\ifnum\minute<10 0\fi\number\minute\the\amorpm}}
\edef\militarytime{\number\hour:\ifnum\minute<10 0\fi\number\minute}

\def\draftlabel#1{{\@bsphack\if@filesw {\let\thepage\relax
   \xdef\@gtempa{\write\@auxout{\string
      \newlabel{#1}{{\@currentlabel}{\thepage}}}}}\@gtempa
   \if@nobreak \ifvmode\nobreak\fi\fi\fi\@esphack}
        \gdef\@eqnlabel{#1}}
\def\@eqnlabel{}
\def\@vacuum{}
\def\draftmarginnote#1{\marginpar{\raggedright\scriptsize\tt#1}}

\def\draft{\oddsidemargin -.5truein
        \def\@oddfoot{\sl preliminary draft \hfil
        \rm\thepage\hfil\sl\today\quad\militarytime}
        \let\@evenfoot\@oddfoot \overfullrule 3pt
        \let\label=\draftlabel
        \let\marginnote=\draftmarginnote
   \def\@eqnnum{(\theequation)\rlap{\kern\marginparsep\tt\@eqnlabel}%
\global\let\@eqnlabel\@vacuum}  }

\def\preprint{\twocolumn\sloppy\flushbottom\parindent 2em
        \leftmargini 2em\leftmarginv .5em\leftmarginvi .5em
        \oddsidemargin -.5in    \evensidemargin -.5in
        \columnsep .4in \footheight 0pt
        \textwidth 10.in        \topmargin  -.4in
        \headheight 12pt \topskip .4in
        \textheight 6.9in \footskip 0pt
        \def\@oddhead{\thepage\hfil\addtocounter{page}{1}\thepage}
        \let\@evenhead\@oddhead \def\@oddfoot{} \def\@evenfoot{} }

\def\numberbysection{\@addtoreset{equation}{section}
        \def\theequation{\thesection.\arabic{equation}}}

\def\underline#1{\relax\ifmmode\@@underline#1\else
        $\@@underline{\hbox{#1}}$\relax\fi}

\def\titlepage{\@restonecolfalse\if@twocolumn\@restonecoltrue\onecolumn
     \else \newpage \fi \thispagestyle{empty}\c@page\z@
        \def\thefootnote{\fnsymbol{footnote}} }

\def\endtitlepage{\if@restonecol\twocolumn \else \newpage \fi
        \def\thefootnote{\arabic{footnote}}
        \setcounter{footnote}{0}}  

\catcode`@=12
\relax

\def\figcap{\section*{Figure Captions\markboth
        {FIGURECAPTIONS}{FIGURECAPTIONS}}\list
        {Figure \arabic{enumi}:\hfill}{\settowidth\labelwidth{Figure
999:}
        \leftmargin\labelwidth
        \advance\leftmargin\labelsep\usecounter{enumi}}}
 \relax
\def\tablecap{\section*{Table Captions\markboth
        {TABLECAPTIONS}{TABLECAPTIONS}}\list
        {Table \arabic{enumi}:\hfill}{\settowidth\labelwidth{Table
999:}
        \leftmargin\labelwidth
        \advance\leftmargin\labelsep\usecounter{enumi}}}
 \relax
\def\reflist{\section*{References\markboth
        {REFLIST}{REFLIST}}\list
        {[\arabic{enumi}]\hfill}{\settowidth\labelwidth{[999]}
        \leftmargin\labelwidth
        \advance\leftmargin\labelsep\usecounter{enumi}}}
 \relax
\makeatletter
\newcounter{pubctr}
\def\publist{\@ifnextchar[{\@publist}{\@@publist}}
\def\@publist[#1]{\list
        {[\arabic{pubctr}]\hfill}{\settowidth\labelwidth{[999]}
        \leftmargin\labelwidth
        \advance\leftmargin\labelsep
        \@nmbrlisttrue\def\@listctr{pubctr}
        \setcounter{pubctr}{#1}\addtocounter{pubctr}{-1}}}
\def\@@publist{\list
        {[\arabic{pubctr}]\hfill}{\settowidth\labelwidth{[999]}
        \leftmargin\labelwidth
        \advance\leftmargin\labelsep
        \@nmbrlisttrue\def\@listctr{pubctr}}}
 \relax
\makeatother
\newskip\humongous \humongous=0pt plus 1000pt minus 1000pt

\newif\ifdtup

\relax

\def\be{\begin{equation}}
\def\ee{\end{equation}}
\def\ba{\begin{eqnarray}}
\def\ea{\end{eqnarray}}

\def\r{\rho}

\def\D{\Delta}

\def\l{\lambda}

\def\no{\noindent}

\def\IR{\relax{\rm I\kern-.18em R}}
\def\II{\relax{\rm 1\kern-.35em1}}

\renewcommand{\theequation}{\thesection.\arabic{equation}}
\csname @addtoreset\endcsname{equation}{section}

\def\IR{\relax{\rm I\kern-.18em R}}
\def\inv{^{\raise.15ex\hbox{${\scriptscriptstyle -}$}\kern-.05em 1}}

\begin{document}

\begin{titlepage}
\begin{center}

\hfill

\vskip 0.1in

{\LARGE Semiclassical correlation functions of Wilson loops \\ and local vertex operators}
\vskip 0.4in

{\bf Rafael Hern\'andez}
 
\vskip 0.1in

Departamento de F\'{\i}sica Te\'orica I\\
Universidad Complutense de Madrid\\
$28040$ Madrid, Spain\\
{\footnotesize{\tt rafael.hernandez@fis.ucm.es}}

\end{center}

\vskip .4in

\centerline{\bf Abstract}
\vskip .2in
\no
We analyze correlation functions of Wilson loop observables and local vertex operators within the strong-coupling regime of the 
AdS/CFT correspondence. When the local operator corresponds to a light string state with finite conserved charges the correlation function 
can be evaluated in the semiclassical approximation of large string tension, where the contribution from the light vertex can be neglected. 
We consider the cases where the Wilson loops are described by two concentric surfaces and the local vertices are the superconformal 
chiral primary scalar or a singlet massive scalar operator.

\noindent

\vskip .4in
\noindent

\end{titlepage}
\vfill
\eject

\def\baselinestretch{1.2}


\baselineskip 20pt


\section{Introduction}

In order to solve a conformal field theory it suffices to determine the spectrum of two and three-point functions 
of primary operators. Higher order correlation functions can be obtained, at least in principle, from these two lower ones 
through the operator product expansion. Conformal symmetry also fixes completely the space-time dependence 
of two and three-point correlation functions. Two-point functions depend only on the spectrum of scaling dimensions of the operators 
of the theory. Three-point functions are constrained by conformal invariance up to the anomalous dimensions and 
some global coefficients, which are the structure constants in the operator product expansion. But in general, except for some protected 
operators, both the scaling dimensions and the structure constants may depend on the coupling constant of the theory. 
Thus although conformal symmetry strongly constrains the operator product expansion complete resolution 
of the spectrum of a conformal field theory turns into a highly involved perturbative problem. 

The AdS/CFT correspondence \cite{Maldacena}-\cite{Gubser} and the uncovering of an integrable 
structure on both sides of the duality (see for instance \cite{intreview} for a comprehensive review on the subject) 
opened a fruitful path towards the resolution of the planar limit of four-dimensional Yang-Mills 
with maximal ${\cal N}=4$ supersymmetry at any value of the gauge coupling constant. 
Within the correspondence the evaluation at strong-coupling of correlation functions 
of single-trace local gauge invariant operators can be performed by inserting closed string vertex operators 
in the path integral for the string partition function. These vertex operators scale exponentially with the energy 
and the quantum conserved charges for the corresponding string states. Therefore when the conserved charges 
are of the order of the string tension the string path integral can be computed in the semiclassical limit of large tension 
through a saddle point approximation. Correlation functions involving {\em heavy} 
states with large conserved charges are then dominated by semiclassical string trajectories. 

The semiclassical approach was first employed in \cite{GKP} to extract the leading order contribution at strong-coupling to the 
cusp anomalous dimension, and has been further explored in the evaluation of diverse two-point functions 
along references~\cite{twopointTseytlin}-\cite{BT}. The extension to correlation functions with two complex conjugate heavy vertex operators 
and one {\em light} string state with fixed conserved charges was recently proposed in~\cite{Zarembo}-\cite{RT} and has been exhaustively 
analyzed for a large variety of heavy vertices and light string states~\cite{H}-\cite{Arnaudov2}.~\footnote{The analysis of the 
more complicated case of three-point correlation functions with three general heavy vertex operators has also been started in \cite{threeh}.} 
The idea is that in the saddle point approximation the leading contribution to the three-point function is coming 
just from the classical string configurations of the vertices with large quantum charges. In the semiclassical limit of 
large string tension the contribution from the light vertex operator can be neglected and the correlation function 
is governed by the classical solutions saturating the two-point function of the heavy operators. Therefore in order 
to find the ratio of the three-point function $\langle V_{H_1}(x_1) V_{H_2}(x_2) V_{L}(x_3) \rangle$ 
and the correlator of the two heavy vertices 
we only need to evaluate the light vertex on the classical configuration,
\be
\frac {\langle V_{H_1}(x_1) V_{H_2}(x_2) V_{L}(x_3) \rangle}{\langle V_{H_1}(x_1) V_{H_2}(x_2) \rangle} 
= \int d^2 \xi \, V_L(x_3)_{\hbox{\tiny{classical}}} \ .
\ee

The semiclassical prescription can also be applied to higher 
$n$-point correlation functions with two heavy vertices and $n-2$ light operators, that turn to be written  
as a product of light vertices evaluated on the classical trajectory determined by the heavy operators. The case 
of four-point functions with two heavy states and two light vertex operators was considered in~\cite{BTfourpoint}. 
As noted above four-point functions can be written in terms of the two lower correlators. This is indeed 
the case for the semiclassical four-point functions in~\cite{BTfourpoint}. But as opposed to 
two and three-point functions, the space-time dependence of four-point and higher order correlation 
functions is not completely constrained by conformal invariance, and in general they will be non-trivial functions 
of the conformal cross ratios of the locations of the operators. In order to understand better the general structure 
of higher order correlation functions it is convenient to analyze some other correlators less constrained by conformal invariance, 
involving for instance other observables in the theory. A natural case we may consider is that of correlation functions with 
Wilson loop observables together with some local operators in the dual gauge theory. Conformal symmetry is not enough 
to fix the space-time dependence and the correlator will depend non-trivially on the position of the vertex operator. 

As in the presence of correlation functions with heavy vertex operators with large quantum charges, correlation 
functions with Wilson loop observables and light vertices are dominated by the minimal surfaces determining the expectation 
value of the Wilson loops at strong-coupling. A semiclassical 
approach has in fact been employed before to evaluate correlation functions with Wilson loops \cite{Berenstein}-\cite{ABT} (see also 
\cite{BPZ,ER} for closely related work). 
In this note we will continue the analysis started in \cite{AT} for correlation functions with two {\em large} Wilson loops 
represented by minimal surfaces ending on some initial and final circles $C_i$ and $C_f$ and one light local vertex operator, 
of the form $\langle W[C_i] W[C_f] V_L(x') \rangle$. At large string tension the contribution from 
the light vertex operator to the stationary surface in the string path integral can again be ignored and it is 
the Wilson loops that dominate the correlation function. The leading contribution to the 
ratio of the three-point function and the correlator of the two Wilson loops
can thus be obtained by evaluating the light vertex operator on the minimal surface that determines the expectation 
value at strong-coupling of the Wilson loop. The normalized correlation function is then given by
\be
\frac {\langle W[C_i] \, W[C_f] \, V_L(x') \rangle}{\langle W[C_i] \, W[C_f] \rangle} = \int d^2 \xi \, V_L(x')_{\hbox{\tiny{Loop}}} \ .
\label{C}
\ee
The case under study in reference \cite{AT} was that of a light dilaton vertex operator. In this note we will extend this proposal to 
explore some other possible choices of light vertices. 
The remaining part of the article is organized as follows. In section 2 we will present an abridged discussion on some relevant features 
of the classical string solutions describing two concentric Wilson loop surfaces that we will consider. In section 3 we will evaluate 
correlation functions with these Wilson loops 
in the cases where the local vertices are the superconformal chiral primary scalar or a singlet massive scalar operator.
In particular we will mostly focus on some degenerate limits of the corresponding minimal surfaces. 
We conclude in section 4 with several general remarks and a discussion on some open problems.


\section{Circular Wilson loop surfaces}

In this section we will briefly review the classical string solution describing a minimal surface that ends 
on two concentric circular Wilson loops at the boundary of $AdS_5$ with angular momentum~$J$ in~$S^5$. 
These surfaces resemble Plateau's problem of soap films in flat spacetime. In this case the minimal surfaces extend 
towards the interior of $AdS_5$. The corresponding solutions were first analyzed in reference \cite{DF}, but here and along this note 
we will follow notation and conventions in \cite{AT}. We will be interested in semiclassical string solutions embedded 
in $AdS_3 \times S^1$, where we will choose coordinates 
\be
ds^2 = z^{-2}(dz^2 + dr^2 + r^2 d \phi^2) + d \varphi^2 \ .
\ee
In these coordinates the minimal surface is described by the ansatz 
\footnote{Here and along this note $\tau$ denotes the euclidean world-sheet time coordinate. The 
minimal surfaces that we are going to consider will thus be embedded in euclidean $AdS_3 \times S^1$.}
\be
z=z(\tau) \ , \quad r=r(\tau) \ , \quad \phi (\sigma) = \sigma \ , \quad \varphi (\tau) = i {\cal J} \tau \ ,
\label{ansatz}
\ee
where ${\cal J} = J/\sqrt{\l}$, together with the boundary conditions
\be
z(\tau_i)=z(\tau_f)= 0 \ , \quad r(\tau_i)=R_i \ , \quad r(\tau_f) = R_f \ ,
\ee
with $\tau_i=0$, and $R_i$ and $R_f$ the radii of the two concentric Wilson loops. In order to find the solution we will need 
the vanishing-energy constraint imposed by the conformal gauge condition 
\be
z^{-2}(\dot{z}^2 + \dot{r}^2 - r^2) = {\cal J}^2 \ , 
\label{constraint1}
\ee
together with the integral of motion
\be
z^{-2}(z \dot{z} + r \dot{r}) = p \ ,
\label{constraint2}
\ee
where $p$ is a constant parameter. If we introduce some new variables $u$ and $v$ through
\be
z = \frac {u e^{v}}{\sqrt{1+u^2}} \ , \quad r = \frac {e^v}{\sqrt{1+u^2}} \ ,
\label{zruv}
\ee
the constraints (\ref{constraint1}) and (\ref{constraint2}) become
\be
\dot{u}^2 = 1 + (1 + {\cal J}^2) u^2 + ({\cal J}^2 - p^2) u^4 \ , \\
\label{uconstraint}
\ee
\be
\dot{v} = \frac {pu^2}{1+u^2} \ .
\label{vconstraint}
\ee
In this note rather than analyzing correlation functions using the most general string solution to these equations 
describing two concentric circular Wilson loops of radii $R_i$ and $R_f$ 
we will mostly consider two degenerate limiting surfaces corresponding to either the case where $p = \pm {\cal J}$ 
or the case where $p=0$ while the angular momentum is kept non-vanishing. We refer the reader to 
reference \cite{DF} for complete details on the more general solutions, and concentrate in what follows on these two degenerate 
minimal surfaces.

In the limit where $p = \pm {\cal J}$ one of the two concentric loops in the general 
ansatz contracts to a point and we are left with a single circular Wilson loop together with an effective heavy vertex 
operator located at the position of the shrunk loop. \footnote{This limit corresponds to the solution first considered in \cite{ZaremboW}.} 
In the $u$ and $v$ variables the degenerate solution becomes
\be
u = \frac {1}{\sqrt{1+p^2}} \sinh \big( \sqrt{1 + p^2} \, \tau \big) \ , \quad 
v = p \tau - \hbox{arctanh} \Big[ \frac {p}{\sqrt{1+p^2}} \tanh \big( \sqrt{1+p^2} \, \tau \big) \Big] \ . 
\label{uv}
\ee
Therefore when $p= + {\cal J}$ we find that $v \rightarrow \infty$ and thus the radius of the outer circle extends to infinity, where 
the effective heavy operator gets located. When we take $p = - {\cal J}$ we find that $v \rightarrow - \infty$ and now it is the inner circle 
that contracts to zero size. The normalized three-point correlation function becomes in both cases that of a single 
circular Wilson loop together with a heavy local operator with large semiclassical 
angular momentum~$J$ and a light vertex operator,
\be
{\cal C}_{W \, V_H V_L} =  
\frac {\langle W[C_f] \, V_H \, V_L \rangle}{\langle W[C_f] \, V_H \rangle} \ .
\ee

In the limit where we set $p=0$ while the angular momentum is kept non-vanishing the constraint~(\ref{constraint2}) implies
\be
z^2 + r^2 = R^2 \ .
\ee
The classical solution reduces to a semi-sphere and the two Wilson loops at the boundary coalesce to a single circle with radius $R$. 
\footnote{This solution is just an extension of the case of 
a minimal surface in $AdS_3$ bounded by a circle constructed in \cite{Berenstein,DGO}, including now an angular momentum \cite{DF}.}
We will take $v=0$ and thus set $R=1$ in the equations below. Now the solution to equation~(\ref{uconstraint}) has two different branches. 
Along the first branch the coordinate $u$ extends from zero at the boundary until infinity. 
Beyond this value the solution continues on the other branch until it reaches the boundary again. 
Writing $\tau$ as a function of $u$ the solution can be expressed in terms of the elliptic and the complete elliptic integrals of the first kind, 
$F(y|m)$ and $K(x)$. For the first branch the solution reads
\be
\tau = \frac {1}{{\cal J}} F \big( \hbox{arctan } u \, {\cal J} \, \big| 1 - 1/{\cal J}^2 \big) \ , 
\label{branch1}
\ee
while for the second branch
\be
\tau = \tau_f - \frac {1}{{\cal J}} F \big( \hbox{arctan } u \, {\cal J} \, \big| 1 - 1/{\cal J}^2 \big) \ .
\label{branch2}
\ee
The time coordinate ranges from $\tau_i=0$ until twice the complete elliptic integral,
\be
\tau_f = \frac {2}{{\cal J}} K \big( 1 - 1/{\cal J}^2 \big) \ .
\ee
The normalized correlation function reduces now to the two-point function of a single circular Wilson 
loop with non-vanishing angular momentum and one light vertex operator, 
\be
{\cal C}_{W \, V_L} =  
\frac {\langle W[C] \, V_L \rangle}{\langle W[C] \rangle} \ .
\ee
This kind of two-point functions were first studied in references \cite{Berenstein}-\cite{Sakaguchi} 
for several different choices of local vertex operators. 


\section{Correlators of Wilson loops and light vertices}

In this section we will follow closely the analysis in \cite{AT} to evaluate the leading order contribution 
in the limit of large string tension to correlation functions of the circular Wilson loop surfaces described 
in the previous section and one light local vertex operator with finite conserved charges. We will explore 
the cases where the light vertices are chosen to be the superconformal primary scalar or a singlet massive 
scalar operator, and find the corresponding normalized correlators for the two limiting solutions discussed 
above where the minimal surfaces degenerate.

\subsection{Superconformal primary scalar operator}

We will first analyze the case where the light vertex operator is taken to be the primary scalar operator. 
The leading contribution in the large string tension expansion to the superconformal primary scalar is 
purely bosonic \cite{Berenstein,Zarembo,RT},
\be
V^{\hbox{\tiny{(primary)}}} (x') = c_{\D_p} \, K_{\D_p}(z;x,x') \, e^{ij \varphi} \, 
\big[ z^{-2} (\partial x_m \bar{\partial}x^m - \partial z \bar{\partial} z ) - \partial \varphi \bar{\partial} \varphi \big] \ ,
\label{Vprimary}
\ee
where $c_{\D_p}$ is the normalization constant of the primary scalar operator, 
$K_{\D_p}(z;x,x')$ is the bulk-to-boundary propagator,  
\be
K_{\D_p}(z;x,x') =  \left[ \frac {z}{z^2+(x-x')^2} \right]^{\Delta_p} \ ,
\ee
and the derivatives are defined as $\partial = \partial_+$ and $\bar{\partial} = \partial_-$. 
The superconformal primary scalar vertex is dual to the BMN operator $\hbox{Tr}Z^j$ and the scaling dimension is just $\D_p = j$. 

As discussed in the introduction, at large string tension the leading order contribution to the normalized correlation function (\ref{C}) is 
dominated by the minimal surface that determines the expectation value of the Wilson loop. 
The correlator can therefore by calculated by evaluating the light vertex operator on the surface (\ref{ansatz}). 
Let us first present the contribution from the propagator in the $u$ and $v$ variables. 
If we parameterize the $x'$ coordinates locating the primary scalar operator 
by $(x_1',x_2')= \rho (\cos \theta, \sin \theta)$, and denote by $h$ the transverse distance in the $(x_3',x_4')$-plane, 
we find \footnote{Conformal symmetry implies that the correlation function of two concentric Wilson loops 
and one local operator should depend only on the radial coordinate on the plane defined by the loops and on the radial coordinate 
on the orthogonal plane (see the appendix in \cite{AT} for a detailed discussion on the general restrictions imposed 
by conformal symmetry on correlators with Wilson loop observables).}
\be
K_{\D_p} = \left[ \frac {u e^v}{\sqrt{1+u^2} \big( e^{2v} + h^2 + \rho^2 \big) - 2 \rho e^v \cos \sigma } \right]^{\D_p} \ ,
\ee
where we have made use of rotational symmetry to remove the dependence on $\theta$ through a shift in $\sigma$. 
The remaining piece in the primary scalar operator (\ref{Vprimary}) can be evaluated 
recalling the conformal constraint (\ref{uconstraint}) and the integral of motion (\ref{vconstraint}). The normalized 
correlation function becomes then
\be
{\cal C} =  2 c_{\D_p} \int_{0}^{\infty} d\tau \int_0^{2\pi} d \sigma \, e^{- {\cal J} j \tau} I(\tau) \, 
\left[ \frac {u e^v}{\sqrt{1+u^2} \big( e^{2v} + h^2 + \rho^2 \big) - 2 \rho \, e^v \cos \sigma} \right]^{\D_p} \ ,
\label{primary}
\ee
where we have defined
\be
I(\tau) = \left[ \frac {p u \mp \sqrt{1 + (1 + {\cal J}^2) u^2 + ({\cal J}^2 - p^2) u^4 }}{1+u^2} \ \right]^2 \ .
\label{I}
\ee
Note that under $p \rightarrow -p$ the two branches of the square root in (\ref{I}) are exchanged.
In general the correlation function (\ref{primary}) will depend on the constant of motion $p$ that parameterizes the minimal surface, 
on the radii of the two concentric Wilson loops and on the parameters $\rho$ and $h$ fixing the position of the light vertex operator 
at the boundary. In what follows we will analyze this dependence in the degenerate limits discussed in the previous section 
where either $p = \pm {\cal J}$ or $p=0$ with non-vanishing angular momentum.
  
\subsubsection{Single Wilson loop and one local operator}
  
In the case where $p = \pm {\cal J}$ either the inner or the outer loops are replaced by an effective local vertex operator 
with angular momentum $J$. In this limit the coordinates $u$ and $v$ are given by equations (\ref{uv}) and the normalized 
correlation function~(\ref{primary}) becomes 
\ba
&& {\cal C}^{(\pm)}_{W \, V_H V_L} =  
2 c_{\D_p} \int_{0}^{\infty} d\tau \int_0^{2\pi} d \sigma \, e^{(\D_p \mp j) p \tau} I(\tau) 
\label{degprimary} \\
&& \times \ \left[ \frac {\sqrt{1+p^2} \tanh g(\tau) - p}{h^2 + \rho^2 + e^{2 p \tau} (1+2p^2) 
- 2 e^{p \tau} \sqrt{1+p^2} \big( \rho \cos \sigma + p e^{p \tau} \sinh g(\tau) \big) \big( \cosh g (\tau) \big)^{-1}} \right]^{\D_p} \ , \nonumber
\ea
where we have set $R_f=1$ and following \cite{AT} we have introduced
\be
g(\tau) = \sqrt{1+p^2} \tau + \hbox{arcsinh} \, p \ .
\ee
Expression (\ref{I}) can be written now as
\be
I(\tau) = \frac {1+p^2}{ \big( (1+p^2 \pm p^2) \cosh g(\tau) - (p \pm p) \sqrt{1+p^2} \sinh g(\tau) \big)^2 } \ . 
\label{Ipm}
\ee
The $\pm$ sign in the correlator (\ref{degprimary}) comes from the choice of degenerating circle in the $p=\pm{\cal J}$ condition 
while in equation (\ref{Ipm}) the upper or the lower signs refer respectively to the positive or negative branches of the square root 
in the general definition of~$I(\tau)$. As in the case of the light dilaton vertex operator analyzed in reference \cite{AT}, 
the correlation functions ${\cal C}^{(+)}$ and ${\cal C}^{(-)}$ are related by an inversion transformation, 
\footnote{An identical transformation property holds also for the more general case of correlation functions involving the non-degenerate 
Wilson loop surfaces studied in \cite{DF}. This is a consequence of the fact that under the inversion symmetry of the AdS metric 
the constraint (\ref{constraint2}) maps into itself with the constant $p$ replaced by $-p$ 
on the right hand side. Therefore minimal surfaces with positive or negative $p$ are related by an inversion transformation.}
\be
{\cal C}^{(\pm)}(p,\D_p,h,\rho) = (h^2 + \rho^2)^{- \D_p} {\cal C}^{(\mp)} \left( -p, \D_p, \frac {h}{h^2 + \rho^2}, \frac {\rho}{h^2 + \rho^2} \right) \ . 
\ee
This is also the behavior in the case of the correlation function with a light singlet scalar vertex operator considered below in this section. 

We have not succeeded in finding a general expression for this correlator in terms of elementary functions. 
However we can still evaluate it in some selected regimes. For instance we may consider the case where the parameter $p$ 
becomes small. In this limit the effective local operator $V_H$ turns into a light vertex, and 
the correlation function reduces to that for a single circular Wilson loop. When the vertex operator is located at the origin where $h=\rho=0$ 
we get
\be
{\cal C}_{W \, V_L} = \frac {4 \pi c_{\D_p}}{\D_p + 1} \ ,
\label{WVprimary}
\ee
which agrees with the result for the correlation function of a circular Wilson loop and one chiral primary operator of charge~$j$ found in \cite{Berenstein}.  
Another interesting limit is that where $p$ becomes large. If we choose $I(\tau)$ along the positive branch 
the leading contribution to expression (\ref{degprimary}) is given by  
\be
{\cal C}_{W \, V_H V_L}^{(\pm)} =
2^{\mp j + 3} |p|^{\mp j + 1}  c_{\D_p} \int_{0}^{\infty} d u \int_0^{2\pi} d \sigma \, 
\frac {u^{\D_p \mp j + 1}}{\big(u^2 + d^2 + 4 \rho \sin^2 (\sigma/2) \big)^{\D_p}} \ ,
\ee
where we have written the integration over the euclidean time coordinate $\tau$ in terms of the variable~$u$. 
The quantity $d^2=h^2+(\rho-1)^2$ is the distance from the insertion point of the primary scalar vertex operator to the Wilson loop.
Evaluating the integrals we obtain
\ba
&& {\cal C}_{W \, V_H V_L}^{(\pm)} = 2^{\mp j + 3} |p|^{\mp j + 1} \pi \, c_{\D_p} \frac {\Gamma[(\D_p \mp j + 2)/2] \Gamma[(\D_p \pm j - 2)/2]}{\Gamma[\D_p]} \nonumber \\ 
\label{Cpmprimarypositive}
&& \times \ d^{\, \mp j - \D_p + 2} \ {}_2F_1 \big( 1/2, (\D_p \pm j - 2)/2, 1, - 4\r/d^2 \, \big) \ .
\ea
If we take the negative branch instead the correlation function becomes
\be
{\cal C}_{W \, V_H V_L}^{(\pm)} =
2^{\mp j - 1} |p|^{\mp j - 3}  c_{\D_p} \int_{0}^{\infty} d u \int_0^{2\pi} d \sigma \, 
\frac {u^{\D_p \mp j - 3}}{\big(u^2 + d^2 + 4 \rho \sin^2 (\sigma/2) \big)^{\D_p}} \ ,
\ee
and upon integration we find
\ba
&& {\cal C}_{W \, V_H V_L}^{(\pm)} = 2^{\mp j - 1} |p|^{\mp j - 3} \pi \, c_{\D_p} \frac {\Gamma[(\D_p \pm j + 2)/2] \Gamma[(\D_p \mp j - 2)/2]}{\Gamma[\D_p]} \nonumber \\ 
\label{Cpmprimarynegative}
&& \times \ d^{\, \mp j - \D_p - 2} \ {}_2F_1 \big( 1/2, (\D_p \pm j + 2)/2, 1, - 4\r/d^2 \, \big) \ .
\ea
Imposing now the marginality condition $\D_p=j$ we find that the correlators $C^{(-)}$ along the positive branch of $I(\tau)$ 
and $C^{(+)}$ along the negative branch are singular. On the contrary the negative choice of branch for the correlator $C^{(-)}$ 
and the positive branch for $C^{(+)}$ provide regular results. 
We note also that when the distance $d^2$ vanishes these correlation functions 
diverge, which is the expected behavior when the light vertex operator approaches the boundary circle of the Wilson loop. 

These correlation functions should be of help to understand the coefficients in the operator product expansion of a circular 
Wilson loop, which is defined as an infinite sum over local operators evaluated at the center of the loop. As discussed in the previous 
section, if we consider for instance the case where $p=-{\cal J}$ it is the inner circle in the general solution that degenerates to a point 
and thus we are evaluating the correlation function of a circular Wilson loop together with a heavy vertex operator located at the origin, 
with effective angular momentum $J$, and a light vertex operator. Then we expect that in the limit where $h=\rho=0$ so that the light vertex 
is at the origin the value of the 
correlator $C^{(-)}$ should be related to an operator product expansion of the circular Wilson loop involving the effective heavy vertex and the 
light chiral primary operator. It would be very interesting to understand better the relation between the above normalized correlation functions 
and the coefficients in the expansion of the circular Wilson loop in terms of local operators, and find out whether a general 
computation can be performed to all orders in the $\alpha'$ expansion of the $AdS_5 \times S^5$ string along the lines of~\cite{ZaremboW}.

\subsubsection{Coincident Wilson loops}
 
When $p=0$ the radii of the loops coincide and we are left with the two-point correlation function of a single Wilson loop 
with angular momentum $J$ and a light primary scalar operator. In order to find the normalized correlator we have to integrate 
the primary scalar vertex over the two branches of the degenerate solution, equations (\ref{branch1}) and (\ref{branch2}). 
We will parameterize again the integration over $\tau$ in terms of the coordinate $u$. Then 
along the first branch we have to integrate $u$ from zero to infinity, and along the second branch we integrate $u$ back to zero. 
We find 
\ba
&& {\cal C}_{W \, V_L} = 2 e^{-K(1-1/{\cal J}^2) j} c_{\D_p} \int_0^{\infty} d u \int_0^{2 \pi} d \sigma \, 
\frac {\big( 1 + (1 + {\cal J}^2) u^2 + {\cal J}^2 u^4 \big) u^{\D_p}}{(1+u^2)^{5/2} \sqrt{1 + {\cal J}^2 u^2}} \label{Ccoincidingprimary} \nonumber \\ 
&& \times \ \frac {\cosh \Big( j \big[ F \big(\hbox{arctan } u{\cal J} \big| 1 - 1/{\cal J}^2 \big) 
- K \big( 1-1/{\cal J}^2 \big) \big] \Big)}{\big( (1+h^2+\rho^2) \sqrt{1+u^2} - 2 \rho \cos \sigma \big)^{\D_p}} \ . 
\ea

If we place the vertex operator at the origin this correlation function can be easily evaluated 
in some special limits. For instance in the case where ${\cal J}$ vanishes the correlator reduces 
to the small $p$ limit in the previous subsection and thus we recover the result (\ref{WVprimary})
for the two-point function of a single circular Wilson loop with no angular momentum and a light 
primary scalar operator \cite{Berenstein}. For arbitrary values of ${\cal J}$ the integral (\ref{Ccoincidingprimary}) 
can be evaluated at large~$j$ by means of the saddle point approximation. The saddle point 
is located at $u=1/\sqrt{{\cal J}}$, and leads to
\be
{\cal C}_{W \, V_L} =  \frac {2 \pi^{3/2} {\cal J}^{1/2} c_{\D_p} }{(1+{\cal J})^{j/2} j^{1/2}} \, e^{-j K( 1-1/{\cal J}^2)} \ .
\ee


\subsection{Singlet massive scalar operator}

We will now consider the case of correlation functions where the light vertex operator is taken to be a singlet massive scalar. 
\footnote{Similar results hold when the light vertex is an operator representing the insertion of string states on the 
leading Regge trajectory with angular momentum $j$ in $S^5$.}
The bosonic piece of the singlet scalar vertex is made out of derivatives of the $S^5$ coordinates~\cite{RTvertex,RT},
\be
V^{\hbox{\tiny{(singlet)}}} (x') = c_{\D_r} \, K_{\D_r}(z;x,x') \big( \partial \varphi  \bar{\partial} \varphi \big)^{r} \ , 
\quad \hbox{with } \ r=2 \ , 4 \ , \ldots 
\label{Vsinglet}
\ee
where now $c_{\D_r} $ is the normalization constant of the singlet scalar operator and 
the scaling dimension is given by $\Delta_r = 2 \sqrt{(r-1)} \lambda^{1/4}$. The value $r=2$ corresponds to a massive string state 
on the first excited level and the corresponding operator in the dual gauge theory is an operator contained within the Konishi multiplet. 
Higher values of $r$ label the remaining levels of order $(r-1)$ in the tower of excited string states. 
  
When we evaluate the singlet scalar vertex operator (\ref{Vsinglet}) in the background of the stationary surface~(\ref{ansatz}) 
the normalized correlator in the $u$ and $v$ variables becomes
\be
{\cal C} = 2 c_{\D_r} \int_{0}^{\infty} d\tau \int_0^{2\pi} d \sigma \, {\cal J}^{2r} \, 
\left[ \frac {u e^v}{\sqrt{1+u^2} \big( e^{2v} + h^2 + \rho^2 \big) - 2 \rho \, e^v \cos \sigma} \right]^{\D_r} \ .
\label{singlet}
\ee
Below we will analyze this correlation function in the two degenerate limiting cases described in the previous section.

\subsubsection{Single Wilson loop and one local operator}
  
Now both the limit where $p=+{\cal J}$ and the outer circle degenerates and the limit 
where $p=-{\cal J}$ so that it is the inner circle that contracts to a point lead to the same correlation function. Using 
solution (\ref{uv}) we find 
\ba
&& {\cal C}_{W \, V_H V_L} =
c_{\D_r} \int_{0}^{\infty} d\tau \int_0^{2\pi} d \sigma \, e^{\D_r  p \tau} p^{2r} 
\label{degsinglet} \\
&& \times \left[ \frac {\sqrt{1+p^2} \tanh g(\tau) - p}{h^2 + \rho^2 + e^{2 p \tau} (1+2p^2) 
- 2 e^{p \tau} \sqrt{1+p^2} \big( \rho \cos \sigma + p e^{p \tau} \sinh g(\tau) \big) \big( \cosh g (\tau) \big)^{-1}} \right]^{\D_r} \, . \nonumber
\ea

As before finding an analytic expression for (\ref{degsinglet}) is a complicated problem. But the correlator  
can again be easily analyzed for small and large values of $p$. 
The small $p$ limit is rather simple because in this case the correlation function vanishes and thus there is no coupling 
between a single circular Wilson loop with no angular momentum and one singlet scalar light operator. 
\no
In the large~$p$ limit the correlator reduces to
\be
{\cal C}_{W \, V_H V_L} = 2 |p|^{2r-1}  c_{\D_r} \int_{0}^{\infty} \frac {d u}{u} \int_0^{2\pi} d \sigma \, 
\left[ \frac {u}{u^2+d^2 + 4 \rho \sin^2 (\sigma/2)} \right]^{\D_r} \ ,
\ee
and evaluating the integrals we now obtain
\be
{\cal C}_{W \, V_H V_L} = 2 |p|^{2r-1} \pi c_{\D_r} \, \frac {\Gamma[\D_r/2]^2}{\Gamma[\D_r]} 
d^{- \D_r} \ {}_2F_1 \big( 1/2, \D_r/2, 1, - 4\r/d^2 \, \big) \ . 
\label{Csinglet} 
\ee
Again when the distance $d^2$ vanishes the light singlet scalar vertex operator approaches the Wilson loop circle at the boundary and 
the correlation function diverges. And when the light vertex is located at the origin we would now expect the correlator to be related to the coefficients 
in the operator product expansion of the Wilson loop in the singlet scalar. 

\subsubsection {Coincident Wilson loops}

In the limit where $p=0$ we need to integrate the singlet scalar vertex operator over the two branches of the solution, 
(\ref{branch1}) and (\ref{branch2}). We find
 \be
{\cal C}_{W \, V_L} =  \int_0^{\infty} d u \int_0^{2 \pi} d \sigma \, 
\frac {2 c_{\D_r} {\cal J}^{2r} u^{\D_r}}{\sqrt{1+u^2}	\sqrt{1 + {\cal J}^2 u^2} \big( (1+h^2+\rho^2) \sqrt{1+u^2} - 2 \rho \cos \sigma \big)^{\D_r} } \ .
\label{coincidingsinglet} 
\ee
Now in the case where the angular momentum vanishes we recover the result for the small~$p$ limit in the previous paragraph. 
However evaluating expression (\ref{coincidingsinglet}) in general for arbitrary values of ${\cal J}$ is a complicated problem 
unless the vertex operator is located at the origin. In this case the above integral can be easily computed and the 
two-point function with non-vanishing angular momentum becomes
\ba
&& {\cal C}_{W \, V_L} =  4 \pi \, c_{\D_r} {\cal J}^{2r-1} \Big[ \pi \, \Gamma[\D_r/2] \
{}_2F_1 \big( 1/2, 1/2, 1 - \D_r/2, 1/{\cal J}^2 \big) \\
&& + {\cal J}^{-\D_r} \: \Gamma[ -\D_r/2] \Gamma[(1+\D_r)/2]^2 \
{}_2F_1 \big( (1+\D_r)/2, (1+\D_r)/2, (2+\D_r)/2, 1/{\cal J}^2 \big) \Big] \ . \nonumber
\ea
The scaling with ${\cal J}^{2r-1}$ appears because the singlet scalar vertex operator is made out of the chiral 
components of the stress tensor for the string sigma model. Therefore when the vertex is evaluated on a classical string 
solution a constant result should be obtained. In fact each partial derivative in the vertex operator provides a factor of ${\cal J}$ 
and upon integration with the bulk-to-bounday propagator an additional factor of ${\cal J}^{-1}$ appears.


\section{Conclusions}

In this note we have analyzed semiclassical correlation functions of Wilson loop observables and one light closed 
string vertex operator within the strong-coupling regime of the AdS/CFT correspondence. We have covered the cases where 
the Wilson loops are described by two concentric surfaces and the local vertices are the superconformal 
chiral primary scalar or a singlet massive scalar operator. We have studied in detail these correlators for some limiting situations 
where the corresponding minimal surfaces degenerate and reduce to either a single circular Wilson loop or a Wilson loop together with an effective 
local operator carrying large angular momentum. In general the correlation functions that we have considered 
exhibit a complicated dependence on the quantum conserved charges labeling the Wilson loop surfaces 
and on the location of the light vertex operator. 

A natural extension of the semiclassical approach in this note is the study of higher order correlation functions 
with a larger amount of light vertex operators. The simplest possibility is the case of  four-point functions 
with two Wilson loop surfaces and two light vertex operators, $\langle W[C_i] \, W[C_f] \, V_{L_1}(x_1) V_{L_2}(x_2) \rangle$. 
At leading order in the limit of large string tension the contribution from the light vertices to the stationary surface dominating 
the string path integral can again be neglected, and the correlation function reduces now to the product 
of the two local operators evaluated on the minimal surface determining the expectation value of the Wilson loops. 
An identical argument holds also for correlators with more than two light vertices.  
The study of this kind of correlation functions and the explicit check of the semiclassical factorization could be of help 
to clarify the general structure of the operator product expansion of local operators. 

Another possible continuation of our analysis is the study of correlation functions involving some other 
Wilson loop observables such as the more general minimal surfaces included in reference \cite{DF}.  
It would also be very interesting to explore the weak-coupling limit of correlation functions of the kind that we have considered. 
The study and comparison of two-point functions on both sides of the AdS/CFT correspondence 
was crucial in order to uncover the integrable structure underlying the duality and magnify our understanding on the spectrum 
of anomalous dimensions. Comparison of three-point functions of non-protected local operators at weak and strong-coupling has in fact been started 
recently in \cite{tailoring}-\cite{Bissi} and has inspired exhaustive spectroscopy of three-point correlators 
\cite{GGGP}, with the motivation to illuminate the possible role played by integrability in the general structure of the complete spectrum 
of the theory. As all order results have been exhibited for Wilson loop observables in the gauge theory~\cite{allorderWilson} 
it could also be expected that the study of more general correlation functions with Wilson loops 
could be of help to clarify the operator product expansion.


\vspace{8 mm}
\centerline{\bf Acknowledgments}

\vspace{2mm}

\no
We would like to thank the organizers of the Nordita Program on ``Exact Results in Gauge-String Dualities''
for warm hospitality and providing a stimulating atmosphere while this note was being completed. 
This work is supported by MICINN through a Ram\'on y Cajal contract and grant FPA2008-04906, 
and by BSCH-UCM through grant GR58/08-910770. 


\end{document}